\documentclass[a4paper,fleqn,usenatbib]{mnras}

\usepackage[T1]{fontenc}
\usepackage{ae,aecompl}

\usepackage{graphicx}	
\usepackage{amsmath}	
\usepackage{amssymb}	
\usepackage{pdflscape}	
\usepackage{verbatim}
\usepackage{color}
\usepackage{pgfplotstable}
\usepackage{booktabs}
\usepackage{rotating}

\title[Red sequence in Abell\,2151]{The build-up of the red sequence in the Hercules cluster}

\author[I. Agulli et al.]{I. Agulli,$^{1,2,3,4}$ \thanks{E-mail: ireagu@iac.es} J. A. L. Aguerri,$^{1,2}$ L. Dominguez Palmero,$^{5,1}$ A. Diaferio$^{3,4}$.\\
$^1$ Instituto de Astrofisica de Canarias, C/ Via Lactea s/n, E-38205 La Laguna, Tenerife, Spain, \\
$^2$ Departamento de Astrofisica, Universidad de La Laguna, E-38206 La Laguna, Tenerife, Spain,\\
$^3$ Dipartimento di Fisica, Universit\`a di Torino, Via P. Giuria 1, I-10125 Torino, Italy\\
$^4$ Istituto Nazionale di Fisica Nucleare (INFN), sezione di Torino, Via P. Giuria 1, I-10125  Torino, Italy\\
$^5$ Isaac Newton Group of Telescopes, Apartado 321, E-38700 Santa Cruz de La Palma, Canary Islands, Spain\\
}

\date{Accepted 2016 May 06. Received 2016 May 05; in original form 2016 April 03}

\pubyear{2016}

\begin{document}
\label{firstpage}
\pagerange{\pageref{firstpage}--\pageref{lastpage}}
\maketitle

\begin{abstract}
We present the study of the colour--magnitude diagram of the cluster Abell 2151 (A\,2151), with a particular focus on the low-mass end. The deep spectroscopy with AF2/WYFFOS@WHT and the caustic method enable us to obtain 360 members within 1.3 $R_{200}$ and absolute magnitude $M_r \lesssim M_r^*+6$. This nearby cluster shows a well defined red sequence up to $M_r \sim - 18.5$; at fainter magnitudes only 36\% of the galaxies lie on the extrapolation of the red sequence. We compare the red sequences of A\,2151 and Abell\,85, which is another nearby cluster with similar spectroscopic data, but with different mass and dynamical state. Both clusters show similar red sequences at the bright end ($M_r \leq -19.5$), whereas large differences appear at the faint end. This result suggests that the reddening of bright galaxies is independent of environment, unlike the dwarf population ($M_r \geq -18.0$).
\end{abstract}




\section{Introduction}\label{intro}
The local Universe shows a clear bimodality in the galaxy populations: galaxies in the colour--colour or colour--magnitude diagrams are located in two different regions: a well--defined red sequence and a blue cloud \citep[e.g.][]{baldry2004}. Using visual and spectra classifications, \cite{strateva2001} showed that the morphology of the two groups is different, the red sequence is dominated by early-type galaxies and the blue cloud by late-type galaxies. \cite{bell2004} found that this bimodality is observed at least out to $z \sim 1$. In addition, \cite{hogg2004} observed a segregation in colour depending on the environment out to similar redshifts. Red and blue galaxies are preferentially located in high- and low-density regions, respectively \citep[e.g.,][]{balogh2004,sanchez2008}. 

We might expect that the formation of the red sequence depends on the mass of the galaxies and/or their environment. The bright end ($M_r \leq -20.0$) of the red sequence seems to be independent of environment \citep[e.g.,][]{hogg2004,depropris2013}, indicating that the reddening of bright galaxies is mainly related to internal processes. The tight red sequence observed for bright galaxies favours scenarios where cluster ellipticals constitute a passive and evolving population formed at high redshift \citep[$z \gtrsim$ 2 -- 3][]{ellis1997,gladders1998,stanford1998} or that massive ellipticals originate from the mergers of massive and metal-rich disc systems \citep{kauffmann1998,delucia2004a,delucia2004b}.

However, at faint magnitudes, the build-up of the red sequence is less clear. Several samples of clusters have observed no evolution of the luminous-to-faint (L/F) ratio with redshift up to $z \sim 1.5$ \citep[][]{lidman2004,andreon2008,crawford2009,depropris2013,andreon2014,cerulo2016}, while a large variation was reported by other studies \citep[][]{delucia2004a,delucia2007,tanaka2005,bildfell2012}.

The redshift variation of the red sequence suggests an evolution of the galaxies from the blue cloud to the red population, originating the so-called Butcher--Oemler effect \citep{butcher1984}. Several mechanisms have ben proposed in order to explain this evolution, in particular, either a combination of dry mergers and quenching of star formation \citep{bell2004,faber2007} or only a late quenching of star formation \citep{cimatti2006,scarlata2007}.

Previous studies on the red sequence evolution focused on galaxies brighter than $M_{r} \sim -19.0$. Deeper spectroscopic surveys are needed to study the physical processes involved in the formation and evolution of the red sequence at fainter magnitudes ($M_{r} > -18.0$). However, the study of clusters at $z \sim 1$ is challenging with present spectroscopic facilities. A possibility would be the analysis and comparison of nearby clusters in different dynamical states.

One of the largest and most massive structures in the Local Universe is the Hercules Supercluster, composed by Abell 2151 (the Hercules cluster and hereafter A\,2151), Abell 2147 and Abell 2152 \citep{chincarini1981,barmby1998}. A\,2151 is a nearby ($z = 0.0367$), irregular and spiral-rich cluster \citep[$\sim 50 $ per cent][]{giovanelli1985} with strong evidence of being in a merging phase and having a large fraction of blue galaxies \citep[e.g.][and references therein]{bird1995,huang1996,dickey1997,cedres2009}. Therefore, A\,2151 appears to be a young cluster, very similar to clusters at higher redshift, but in the nearby Universe, allowing deep spectroscopic observations of its central region down to $\sim M_r^* + 6$. We emphasize that this faint magnitude limit probes the dwarf galaxy population because it corresponds to a mass M$_* \sim 3.7\times10^8$\,M$_{\odot}$ for a red sequence member. Therefore, the young dynamical state, the variety of local densities, and the large fraction of blue galaxies make A\,2151 an ideal environment to study the build-up of the red sequence at faint magnitudes.

We will briefly discuss the data set in Section 2, we present the results in Section 3, and the discussion and conclusion in Section 4. Throughout this work, we use the cosmological parameters $H_0 = 75 \; \mathrm{km} \, \mathrm{s}^{-1} \mathrm{Mpc}^{-1}$, $\Omega _m = 0.3$ and $\Omega _{\Lambda} = 0.7$. 

\section{The observational data of A\,2151}
\subsection{Deep AF2/WYFFOS spectroscopy}
\begin{figure}
\centering
\includegraphics[width=1\linewidth]{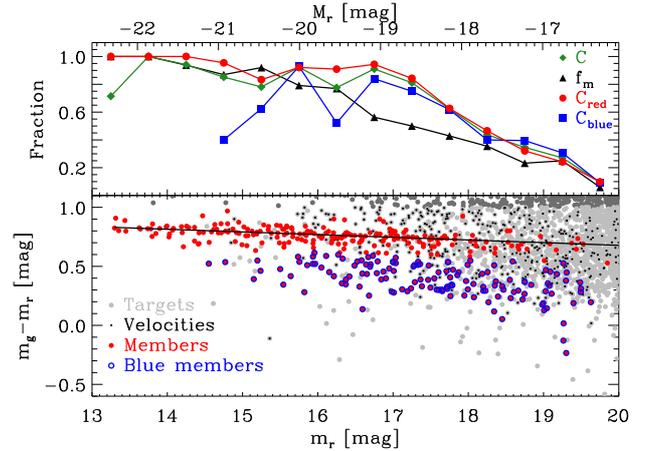}
  \caption{Lower panel: colour--magnitude diagram of the galaxies in the direction of A\,2151. Dark grey dots are the galaxies excluded from target with the colour-cut. Light grey dots are the target galaxies and black points are the velocities obtained. Red and blue symbols show red and blue cluster members, respectively. The solid line represents the red sequence of the cluster. Upper panel: spectroscopic completeness  ($C$, green diamonds), red ($C_\text{red}$, red dots) and blue ($C_\text{blue}$, blue squares) spectroscopic completeness, and cluster member fraction ($f_\text{m}$, black triangles) as a function of $r$-band magnitude.}
\label{cmd}
\end{figure}

Our parent photometric catalogue contains all galaxies brighter than $m_r = 20.5$ mag\footnote{The apparent magnitudes used here are the model SDSS-DR9 $r$-band magnitudes corrected for extinction.} from the SDSS-DR9 \citep[][]{ahn2012}, and within 45 arcmin from the cluster mass centre\footnote{$\alpha$ (J2000): $16^\text{h} \, 05^\text{m} \, 26^\text{s}$, $\delta$ (J2000): $17^{\circ} \, 44' \, 55"$.} \citep{sanchez2005}. We present the target selection in the colour--magnitude diagram (see Fig. \ref{cmd}). We select those galaxies bluer than $m_g -m_r \leq 1.0$, with apparent magnitude brighter than 20, and with no spectroscopic measurements available in literature. This colour cut should match the colour distribution of galaxies in the nearby Universe and minimize the background source contamination \citep[see][]{hogg2004,rines2008}. 

We observed these objects during three nights plus 8 h of service time at the \textit{William Herschel Telescope} (\textit{WHT}) with the fibre spectrograph AutoFib2/WYFFOS using the R158B grism ($R\,=\,280$). The observations were designed to maximize the number of targets within 20 arcmin -- where the instrument response is optimal, eight pointings with  $\sim 90$ galaxies each. We reached a signal to noise higher than 5 for the faintest galaxies with three exposures of 1800 s per pointing. We obtained 738 spectra and  reduced them with the version 2.25 of the instrument pipeline \citep{dominguez2014}.

\subsection{Velocity and member catalogues}
We used the \textit{rvsao.xcsao} \textit{IRAF} task \citep{kurtz1992} to determine the recessional velocities of the observed galaxies. This task cross-correlates a template spectrum library \citep[in this work,][]{kennicutt1992} with the observed galaxy spectrum. For a full description of the task and the technique we refer the reader to \cite{kurtz1992}. We could determine 453 recessional velocities, the remaining spectra were too noisy to rely on the results of the cross-correlation. In fact,  these data enabled us to determine that a mean surface brightness of 23 mag arcsec$^{-2}$ is the instrument limit for our observations. In general, the formal errors on the velocities provided by this task are smaller than the realistic uncertainties: we estimated reliable errors by comparing the results for the same objects observed in different pointings. The root mean square of the differences in measured velocities for the 57 objects with repeated observation is 175 km\,s$^{-1}$. We also had between 2 and 4 galaxies per pointing with spectroscopic information in SDSS-DR9. The differences between our measured velocities and SDSS-DR9 ones are always smaller than 100 km\,s$^{-1}$. 

The redshifts from the literature (SDSS-DR9 and NED catalogues), together with our new data, result in a total number of 799 galaxy velocities within a radius of 0.94 Mpc from the cluster centre; 362 of them are new velocities with $-18.3 \leq M_r \leq -16.0$. Fig. \ref{cmd} presents the completeness of the spectroscopic data set defined as $C = N_z / N_\text{phot}$, with $N_z$ being the number of measured redshifts and $N_\text{phot}$ the number of photometric targets. $C$ is higher than 80\% for $M_r \lesssim -18.5$, and $\sim 30 \% $ at $M_r  \sim -17$. 

\begin{figure}
\centering
\includegraphics[width=1\linewidth]{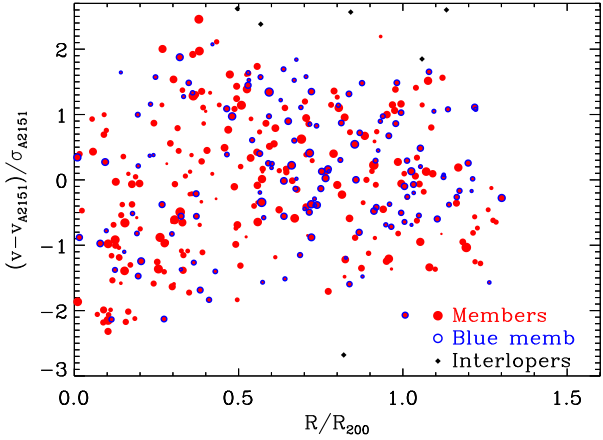}
  \caption{Line-of-sight velocity -- projected clustercentric distance plane for the A2151 members (red dots) and the interlopers (black diamonds). The dot dimensions scale with the magnitude: bright (faint) -- big (small) dots. Blue circles identify blue members.}
\label{caust}
\end{figure}

The cluster membership was determined with the caustic method \citep{diaferio1997,diaferio1999,serra2011}. This technique estimates the escape velocity and the mass profile of clusters in both the virial and infall regions, without any assumption of dynamical equilibrium. A by-product of this technique is the member identification. The members were identified by the binary tree, that gives an interloper contamination of only 3\% within $R_{200}$ \citep{serra2013}. We obtained $v_\text{c} = 10885$ km/s, $\sigma_\text{c}=704$ km/s and 360 members. Fig. \ref{caust} shows the line-of-sight velocity -- projected clustercentric distance plane of A\,2151, where the technique evaluates the caustic amplitudes and the radial mass profile. From this profile, we estimated $M_{200} = 4.00 \times10^{14}$\,M$_\odot$ and $R_{200} = 1.45 $ Mpc. The member fraction is presented in Fig. \ref{cmd} and is defined as $f_\text{m} = N_\text{m} / N_z$, being $N_\text{m}$ the number of members. This fraction strongly depends on the luminosity. 

\section{Results}
\subsection{The colour--magnitude diagram}
We performed a linear fit of the red sequence of A\,2151 using all the members brighter than $m_r = 17$ and with $(m_g -m_r) > 0.6$. The best fit is $(m_g -m_r)_\text{RS} = -0.022m_r + 1.132$ and $\sigma_\text{RS} = 0.053$. The galaxies with $(m_g -m_r)$ smaller or greater  than $ (m_g -m_r)_\text{RS} - 3 \sigma_\text{RS}$ are blue and red galaxies, respectively. Fig. \ref{cmd} shows the two populations and the fitted red sequence. A tight red sequence is visible up to $M_r \sim -18.5$; the scatter increases at fainter magnitudes. In particular, only 36 \% of the galaxies fainter than $M_r =-18$ are classified as red. 

The absence of red dwarfs\footnote{Galaxies are defined as dwarf or faint (F) if $M_r \geq -18$, and as bright or luminous (L) if $M_r \leq -20$.} is not an observational bias. In fact, we measured the recessional velocities of red galaxies, as a visual inspection of Fig. \ref{cmd} shows, but these galaxies result to be background objects. To better quantify the absence of any bias, we defined the red and blue completenesses, $C_i = N_{z,i} / N_{\text{phot},i}$ where $_i = _\text{\{red,blue\}}$ and the numbers of the two populations in the target sample are based on the definitions mentioned above. Using the red sequence of the cluster to separate the two populations is appropriate because we thus estimate the possible loss of red members of the cluster. As Fig. \ref{cmd} shows, the two fractions have similar trends to the global completeness, $C$, at the low-mass end. Therefore, the deficiency of red dwarfs in A\,2151 is real. Moreover, among the red dwarfs within 5 $\sigma_\text{c}$ from $v_\text{c}$, the caustic method only removes four galaxies from the member list which do not affect the results.

\begin{figure*}
\centering
\includegraphics[width=1\linewidth]{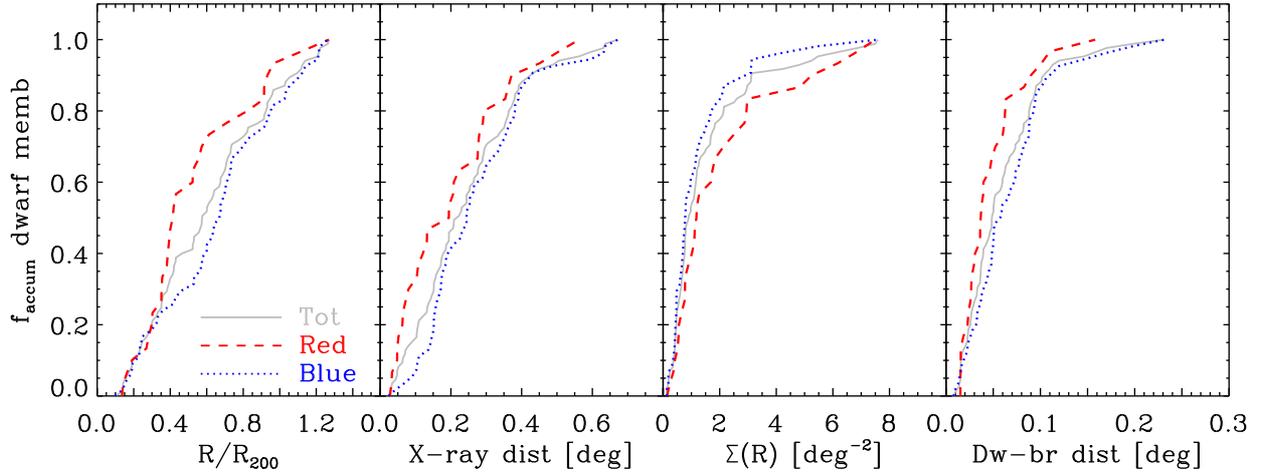}
  \caption{The cumulative distribution functions for the dwarf members ($M_r \geq -18$) of A\,2151. Grey solid lines are for the whole sample, red dashed lines for the red sample and blue dotted lines for the blue sample. The first panel from the left is for the clustercentric distance, the second for the distance of each galaxy to the nearest X-ray peak, the third for the local density and the fourth for the distance of each dwarf to the nearest bright galaxy ($M_r \leq -20$).}
\label{accum_dist}
\end{figure*}

\subsection{Red and blue populations in Abell 2151}
To understand the deficiency of red dwarfs in A\,2151, we studied the dependence of the two populations, red and blue, on different variables. In particular, we used the clustercentric distance, the distance of each member to the nearest X-ray peak, the local density and the distance of each dwarf to the nearest bright galaxy. We consider the cluster centre provided by the caustic technique. This centre is only 80 kpc away from the mass centre estimated by \citet{sanchez2005}. We considered the four X-ray peaks analysed by \cite{bird1995} of the X-ray gas distribution measured by \textit{ROSAT}. We evaluated the local density of the galaxy distribution with the 2D-DEDICA algorithm by \cite{pisani1993}. The global cluster environment is related to the clustercentric distance and the distance to the nearest X-ray peaks, whereas the local environment is connected to the local density and the distance between dwarf and bright galaxies. 

Fig. \ref{accum_dist} shows the cumulative distribution functions of the total, red and blue dwarf galaxies for the physical parameters mentioned above. A visual inspection suggests differences between the red and blue populations for all variables. In general, compared to blue galaxies, red dwarf galaxies are located closer to the cluster centre, to the X-ray peaks and to bright galaxies, and their local densities are larger. However, not all these differences are significant. The Kolmogorov--Smirnov test at $> 99$ per cent C.L. shows that the red and blue populations are statistically different only as far as the clustercentric and the X-ray peak distances are concerned. On the contrary, we can not exclude the same parent distribution when we consider the remaining two physical parameters. These results indicate that the distribution of the red and blue dwarfs are more sensitive to the global, rather than local, cluster properties.

\subsection{Comparison with Abell 85}
\begin{figure}
\centering
\includegraphics[width=1\linewidth]{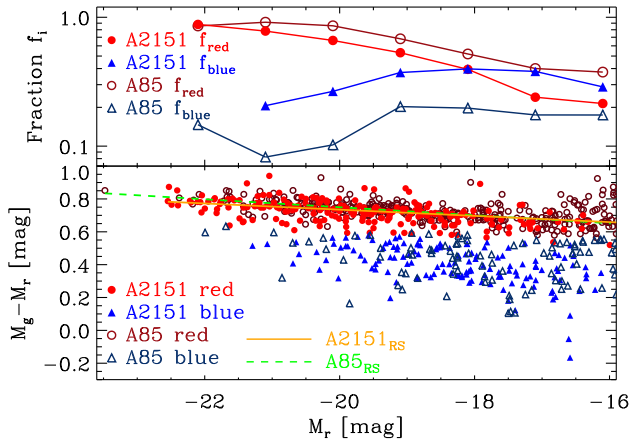}
  \caption{Upper panel: red (red circles) and blue (blue triangles) fractions for A\,2151 (solid symbols) and for A\,85 (open symbols). Lower panel: the colour--magnitude diagram of the two clusters with the same colour and symbol code as the upper panel. The straight lines represent the red sequence of A\,2151 (solid orange) and A\,85 (dashed green).}
\label{cmd_a2151_a85}
\end{figure}

Fig. \ref{cmd_a2151_a85} presents the comparison between the  red sequences of A\,2151 and Abell~85 (A\,85, hereafter), which is another nearby and massive cluster, with similar spectroscopic information \citep{agulli2014,agulli2016}. For A\,85, we have 460 spectroscopically confirmed members. In order to compare the samples, we calculated the absolute magnitude applying the \textit{K}--correction given by \cite{chilingarian2012}. Both red sequences have been fitted with the members in the same magnitude and colour ranges. The zero--point and the slope of the two best models are compatible within the errors. However, the faint end and the relative fraction of red and blue dwarfs are clearly different. Indeed, in A\,2151 the blue populations dominates at $M_r \geq -18$, while in A\,85 the fraction of red galaxies is larger in the full magnitude range. 

\section{Discussion and conclusion}
Several investigations find that the red-sequence slope is independent of the environment \citep[e.g.][]{delucia2004b,delucia2007,tanaka2005}. This result is confirmed by A\,2151 and A\,85 whose slopes are comparable, even if these two clusters have different dynamical state and local environment. Indeed, A\,2151 shows ongoing major merging events, has substructures and a complex X-ray map \citep[e.g.,][]{bird1995,huang1996,cedres2009}, while A\,85 appears to be almost virialized, with small substructures and a smoother X-ray map \citep[e.g.,][]{ramella2007,boue2008}. Although a direct comparison with other slopes presented in the literature is difficult due to the different filters and the possible biases, our result is in general agreement with many studies \citep[e.g.,][]{delucia2004b,delucia2007,tanaka2005,cerulo2016}.

A consequence of this red-sequence invariance is that the reddening of the bright galaxies is more likely due to internal causes at least earlier than redshift $\sim 1$, which is the redshift coverage of previous studies. The bright end also is independent of the definition of environment: in fact, the result for the bright galaxies we obtain here, where we compare two clusters in different dynamical states, is similar to the results of \cite{delucia2007} who separate massive and small clusters based on the velocity dispersion, and the results of \cite{tanaka2005} who used the local density to define field, groups and clusters.

The build-up of the faint part of the red sequence, instead, is less clear: for example, \cite{delucia2007} find a redshift evolution of the L/F ratio (the ratio between the numbers of bright to faint galaxies) that is not confirmed by \cite{andreon2008} and \cite{cerulo2016}. However, the magnitude limit of these spectroscopic studies is quite bright: generally is $M_r \leq -19$, that is one magnitude brighter than the usually assumed largest luminosity of dwarf galaxies ($M_r = -18$). The data set analysed here goes two magnitude fainter than the dwarf limit. The same limit applies to the sample of A\,85 used for the comparison. Our deeper limit makes the direct comparison with the literature results more difficult. By defining dwarf or faint the galaxies fainter than $M_r = -18$, we observe a large difference between the two clusters: the dwarf population of A\,85 is dominated by red galaxies, whereas the A\,2151 dwarf population is dominated by blue galaxies; this results is confirmed by the ratio L/F$= 0.46$ of the red population for A\,85 and L/F$=2.6$ for A\,2151. Because the main differences between these two clusters are their mass and dynamical state, our result suggests that the build-up of the faint end  is related to these properties. 

We find that the differences between the cumulative distribution functions of two global cluster parameters, the clustercentric and nearest X-ray peak distances of the individual galaxies, of the red and blue dwarf populations of A\,2151 are statistically significant. Our result suggests that either dwarf galaxies have not lived long enough in the cluster hostile environment or the environmental processes are not strong enough to remove the gas from the dwarfs and to transform them from active star forming faint galaxies to passive red dwarfs. This transformation has already started for those galaxies embedded within the peaks of the X-ray emission. Our findings agree with \cite{raichoor2014} who find that  the faction of blue galaxies within a cluster only depends on the stellar mass of the galaxies and their clustercentric distances.

The luminosity function is a powerful observable to analyse the characteristics of different galaxy populations and the effect of the environment on them. A detailed analysis of the luminosity function of A\,2151 will be presented in a forthcoming paper (Agulli et al., in preparation).

In conclusion, we find that the slope and the zero--point of the red sequence appears to be well defined, unlike its trend at faint magnitudes, in agreement with the analysis of \cite{delucia2007}. Our comparison between A\,2151 and A\,85 presented here suggest a strong correlation between this behaviour and the dynamical and accretion histories of the clusters. 

\section*{Acknowledgements.}
We acknowledge the anonymous referee for the useful comments that helped us to improve the Letter. We would like to acknowledge Ana Laura Serra and Heng Yu for the helpful discussion on the results from the caustic method. This work has been partially funded by the MINECO (grant AYA2013-43188-P). IA and AD acknowledge partial support from the INFN grant InDark. This research has made use of the Ninth Data Release of SDSS, and of the NASA/IPAC Extragalactic Database which is operated by the Jet Propulsion Laboratory, California Institute of Technology, under contract with the National Aeronautics and Space Administration. The \textit{WHT} and its service programme are operated on the island of La Palma by the Isaac Newton Group in the Spanish Observatorio del Roque de los Muchachos of the Instituto de Astrof\'isica de Canarias.





\begin{thebibliography}{99}
%
\bibitem[\protect\citeauthoryear{Ahn et al.}{2012}]{ahn2012} Ahn C.~P., et al., 2012, ApJS, 203, 21 
%
\bibitem[\protect\citeauthoryear{Agulli et al.}{2014}]{agulli2014} Agulli I., Aguerri J.~A.~L., S{\'a}nchez-Janssen R., Barrena R., Diaferio A., Serra A.~L., M{\'e}ndez-Abreu J., 2014, MNRAS, 444, L34 
%
\bibitem[\protect\citeauthoryear{Agulli et al.}{2016}]{agulli2016} Agulli I., Aguerri J.~A.~L., S{\'a}nchez-Janssen R., Dalla Vecchia C., Diaferio A., Barrena R., Dominguez Palmero L., Yu H., 2016, MNRAS, 458, 1590 
%
\bibitem[\protect\citeauthoryear{Andreon et al.}{2008}]{andreon2008} Andreon S., Puddu E., de Propris R., Cuillandre J.-C., 2008, MNRAS, 385, 979 
%
\bibitem[\protect\citeauthoryear{Andreon et al.}{2014}]{andreon2014} Andreon S., Newman A.~B., Trinchieri G., Raichoor A., Ellis R.~S., Treu T., 2014, A\&A, 565, A120 
%
\bibitem[\protect\citeauthoryear{Baldry et al.}{2004}]{baldry2004} Baldry I.~K., Glazebrook K., Brinkmann J., Ivezi{\'c} {\v Z}., Lupton R.~H., Nichol R.~C., Szalay A.~S., 2004, ApJ, 600, 681 
%
\bibitem[\protect\citeauthoryear{Balogh et al.}{2004}]{balogh2004} Balogh M.~L., Baldry I.~K., Nichol R., Miller C., Bower R., Glazebrook K., 2004, ApJ, 615, L101 
%
%
\bibitem[\protect\citeauthoryear{Barmby \& Huchra}{1998}]{barmby1998} Barmby P., Huchra J.~P., 1998, AJ, 115, 6 
%
\bibitem[\protect\citeauthoryear{Bell et al.}{2004}]{bell2004} Bell E.~F., et al., 2004, ApJ, 608, 752 
%
\bibitem[\protect\citeauthoryear{Bildfell et al.}{2012}]{bildfell2012} Bildfell C., et al., 2012, MNRAS, 425, 204 
%
\bibitem[\protect\citeauthoryear{Bird, Davis, \& Beers}{1995}]{bird1995} Bird C.~M., Davis D.~S., Beers T.~C., 1995, AJ, 109, 920 
%
\bibitem[\protect\citeauthoryear{Bou{\'e} et al.}{2008}]{boue2008} Bou{\'e} G., Durret F., Adami C., Mamon G.~A., Ilbert O., Cayatte V., 2008, A\&A, 489, 11 
%
\bibitem[\protect\citeauthoryear{Butcher \& Oemler}{1984}]{butcher1984} Butcher H., Oemler A., Jr., 1984, ApJ, 285, 426 
%
\bibitem[\protect\citeauthoryear{Cedr{\'e}s et al.}{2009}]{cedres2009} Cedr{\'e}s B., Iglesias-P{\'a}ramo J., V{\'{\i}}lchez J.~M., Reverte D., Petropoulou V., Hern{\'a}ndez-Fern{\'a}ndez J., 2009, AJ, 138, 873
%
\bibitem[\protect\citeauthoryear{Cerulo et al.}{2016}]{cerulo2016} Cerulo P., et al., 2016, MNRAS, 457, 2209 
%
\bibitem[\protect\citeauthoryear{Chilingarian \& Zolotukhin}{2012}]{chilingarian2012} Chilingarian I.~V., Zolotukhin I.~Y., 2012, MNRAS, 419, 1727 
%
\bibitem[\protect\citeauthoryear{Chincarini, Thompson, \& Rood}{1981}]{chincarini1981} Chincarini G., Thompson L.~A., Rood H.~J., 1981, ApJ, 249, L47 
%
\bibitem[\protect\citeauthoryear{Cimatti, Daddi, \& Renzini}{2006}]{cimatti2006} Cimatti A., Daddi E., Renzini A., 2006, A\&A, 453, L29 
%
\bibitem[\protect\citeauthoryear{Crawford, Bershady, \& Hoessel}{2009}]{crawford2009} Crawford S.~M., Bershady M.~A., Hoessel J.~G., 2009, ApJ, 690, 1158 
%
\bibitem[\protect\citeauthoryear{De Lucia, Kauffmann, \& White}{2004a}]{delucia2004a} De Lucia G., Kauffmann G., White S.~D.~M., 2004a, MNRAS, 349, 1101 
%
\bibitem[\protect\citeauthoryear{De Lucia et al.}{2004b}]{delucia2004b} De Lucia G., et al., 2004b, ApJ, 610, L77 
%
\bibitem[\protect\citeauthoryear{De Lucia et al.}{2007}]{delucia2007} De Lucia G., et al., 2007, MNRAS, 374, 809 
%
%
\bibitem[\protect\citeauthoryear{De Propris, Phillipps, \& Bremer}{2013}]{depropris2013} De Propris R., Phillipps S., Bremer M.~N., 2013, MNRAS, 434, 3469 
%
\bibitem[\protect\citeauthoryear{Diaferio \& Geller}{1997}]{diaferio1997} Diaferio A., Geller M.~J., 1997, ApJ, 481, 633 
%
\bibitem[\protect\citeauthoryear{Diaferio}{1999}]{diaferio1999} Diaferio A., 1999, MNRAS, 309, 610 
%
\bibitem[\protect\citeauthoryear{Dickey}{1997}]{dickey1997} Dickey J.~M., 1997, AJ, 113, 1939 
%
\bibitem[\protect\citeauthoryear{Dom{\'{\i}}nquez Palmero et al.}{2014}]{dominguez2014} Dom{\'{\i}}nquez Palmero L., Jackson R., Molaeinezhad A., Fari{\~n}a C., Balcells M., Benn C.~R., 2014, SPIE, 9149, 91492J 
%
\bibitem[\protect\citeauthoryear{Ellis et al.}{1997}]{ellis1997} Ellis R.~S., Smail I., Dressler A., Couch W.~J., Oemler A., Jr., Butcher H., Sharples R.~M., 1997, ApJ, 483, 582 
%
\bibitem[\protect\citeauthoryear{Faber et al.}{2007}]{faber2007} Faber S.~M., et al., 2007, ApJ, 665, 265
%
\bibitem[\protect\citeauthoryear{Giovanelli \& Haynes}{1985}]{giovanelli1985} Giovanelli R., Haynes M.~P., 1985, ApJ, 292, 404 
%
\bibitem[\protect\citeauthoryear{Gladders et al.}{1998}]{gladders1998} Gladders M.~D., L{\'o}pez-Cruz O., Yee H.~K.~C., Kodama T., 1998, ApJ, 501, 571 
%
\bibitem[\protect\citeauthoryear{Hogg et al.}{2004}]{hogg2004} Hogg D.~W., et al., 2004, ApJ, 601, L29 
%
\bibitem[\protect\citeauthoryear{Huang \& Sarazin}{1996}]{huang1996} Huang Z., Sarazin C.~L., 1996, ApJ, 461, 622 
%
\bibitem[\protect\citeauthoryear{Kauffmann \& Charlot}{1998}]{kauffmann1998} Kauffmann G., Charlot S., 1998, MNRAS, 294, 705 
%
\bibitem[\protect\citeauthoryear{Kennicutt}{1992}]{kennicutt1992} Kennicutt R.~C., Jr., 1992, ApJ, 388, 310 
%
\bibitem[\protect\citeauthoryear{Kurtz et al.}{1992}]{kurtz1992} Kurtz M.~J., Mink D.~J., Wyatt W.~F., Fabricant D.~G., Torres G., Kriss G.~A., Tonry J.~L., 1992, ASPC, 25, 432 
%
\bibitem[\protect\citeauthoryear{Lidman et al.}{2004}]{lidman2004} Lidman C., Rosati P., Demarco R., Nonino M., Mainieri V., Stanford S.~A., Toft S., 2004, A\&A, 416, 829 
%
\bibitem[\protect\citeauthoryear{Pisani}{1993}]{pisani1993} Pisani A., 1993, MNRAS, 265, 706 
%
\bibitem[\protect\citeauthoryear{Raichoor \& Andreon}{2014}]{raichoor2014} Raichoor A., Andreon S., 2014, A\&A, 570, A123 
%
\bibitem[\protect\citeauthoryear{Ramella et al.}{2007}]{ramella2007} Ramella M., et al., 2007, A\&A, 470, 39
%
\bibitem[\protect\citeauthoryear{Rines \& Geller}{2008}]{rines2008} Rines K., Geller M.~J., 2008, AJ, 135, 1837 
%
\bibitem[\protect\citeauthoryear{S{\'a}nchez-Janssen et al.}{2005}]{sanchez2005} S{\'a}nchez-Janssen R., Iglesias-P{\'a}ramo J., Mu{\~n}oz-Tu{\~n}{\'o}n C., Aguerri J.~A.~L., V{\'{\i}}lchez J.~M., 2005, A\&A, 434, 521
%
\bibitem[\protect\citeauthoryear{S{\'a}nchez-Janssen, Aguerri, \& Mu{\~n}oz-Tu{\~n}{\'o}n}{2008}]{sanchez2008} S{\'a}nchez-Janssen R., Aguerri J.~A.~L., Mu{\~n}oz-Tu{\~n}{\'o}n C., 2008, ApJ, 679, L77 
%
\bibitem[\protect\citeauthoryear{Scarlata et al.}{2007}]{scarlata2007} Scarlata C., et al., 2007, ApJS, 172, 494 
%
\bibitem[\protect\citeauthoryear{Serra et al.}{2011}]{serra2011} Serra A.~L., Diaferio A., Murante G., Borgani S., 2011, MNRAS, 412, 800 
%
\bibitem[\protect\citeauthoryear{Serra \& Diaferio}{2013}]{serra2013} Serra A.~L., Diaferio A., 2013, ApJ, 768, 116 
%
\bibitem[\protect\citeauthoryear{Stanford, Eisenhardt, \& Dickinson}{1998}]{stanford1998} Stanford S.~A., Eisenhardt P.~R., Dickinson M., 1998, ApJ, 492, 461 
%
\bibitem[\protect\citeauthoryear{Strateva et al.}{2001}]{strateva2001} Strateva I., et al., 2001, AJ, 122, 1861 
%
\bibitem[\protect\citeauthoryear{Tanaka et al.}{2005}]{tanaka2005} Tanaka M., Kodama T., Arimoto N., Okamura S., Umetsu K., Shimasaku K., Tanaka I., Yamada T., 2005, MNRAS, 362, 268 
%
%
%
\end{thebibliography}



\appendix


\bsp	
\label{lastpage}
\end{document}